\documentclass[a4paper,aps,prd,10pt,preprintnumbers,showpacs,superscriptaddress,nofootinbib,amsmath,amssymb,floatfix
,twocolumn
]{revtex4-1}

\usepackage{enumitem}
\usepackage{graphicx}
\usepackage[utf8]{inputenc}
\usepackage[T1]{fontenc}
\usepackage{cmap}
\usepackage{graphicx}
\usepackage{dcolumn}
\usepackage{bm}
\usepackage{bbm}
\usepackage{amsfonts}
\usepackage{amsmath}
\usepackage{enumitem}
\usepackage{lipsum}
\usepackage[bottom]{footmisc}
\usepackage{cmap}
\usepackage{subcaption}
\usepackage{braket}
\usepackage{braket}
\usepackage[table]{xcolor}
\usepackage{amsmath}
\usepackage{hyperref}
\usepackage{cleveref}
\usepackage{fmtcount}
\usepackage{dblfnote}
\usepackage{float}
\usepackage{amssymb}
\usepackage{natbib}
\usepackage{blindtext}

\begin{document}
\title{Novel topological subclass in Ho\u{r}ava-Lifshitz black holes}
\author{Hao Chen}
\email{haochen1249@yeah.net}
\affiliation{School of Physics and Electronic Science, Zunyi Normal University, Zunyi 563006,PR China}
\author{Meng-Yao Zhang}
\affiliation{College of Computer and Information Engineering, Guizhou University of Commerce, Guiyang, 550014, China}
\author{Hassan Hassanabadi}
\affiliation{Departamento de F\'{\i}sica Te\'orica, At\'omica y Optica and Laboratory for Disruptive \\ Interdisciplinary Science (LaDIS), Universidad de Valladolid, 47011 Valladolid, Spain}
\affiliation{Department   of   Physics,   University   of   Hradec   Kr\'{a}lov\'{e}, Rokitansk\'{e}ho   62, 500   03   Hradec   Kr\'{a}lov\'{e},   Czechia}
\affiliation{Department of Physics and Electronics,  Khazar University, 41 Mahsati Str, AZ1096, Baku, Azerbaijan}
\author{Qihong Huang}
\affiliation{School of Physics and Electronic Science, Zunyi Normal University, Zunyi 563006,PR China}
\author{Zheng-Wen Long}
\affiliation{College of Physics, Guizhou University, Guiyang, Guizhou 550025, People's Republic of China}


\begin{abstract}
This work explores the universal classification of thermodynamic topology for charged static black holes within the $z=3$ Ho\u{r}ava-Lifshitz gravity theory, considering both canonical and grand canonical ensembles. We introduce a new topological subclass, denoted as $\ddot{W}^{1-}$. This finding expands the existing topological classification, going beyond the five previously defined classes and their respective subclasses. The $\ddot{W}^{1-}$ subclass presents a distinct and previously unobserved stability profile: In the low-temperature regime, an unstable small black hole appears in the phase space, whereas, while in the high temperature regime, two unstable small black holes exist together with a stable large black hole. Our study underscores the dependence of charged black hole stability on the selection of the ensemble. These results contribute to refining and expanding the topological framework in black hole thermodynamics, providing key perspectives on the underlying nature of black holes and gravity.
\end{abstract}
\maketitle

\section{Introduction}\label{Sec1}
The field of black hole thermodynamics has long attracted significant attention, offering essential insights into the fundamental links between quantum information, statistical mechanics, and geometric properties \cite{qq1,qq2}. Phase transitions are widespread in thermodynamic systems. By taking AdS space as the background and studying Schwarzschild black holes as the thermodynamic system, it has been discovered that the critical temperature of a black hole determines its stability, a process known as the Hawking-Page phase transition \cite{ch01}, which has major consequences for gauge field theory, particularly through the AdS/CFT duality. The fundamental principle behind the comprehensive investigation of black hole chemistry is to view the cosmological constant as a thermodynamic quantity (pressure), whose conjugate variable is naturally associated with the black hole volume. This approach reveals the novel phase transitions and intricate phase structures \cite{qq3,qq4,qq5,qq6,ka1}.

Despite notable advances in black hole thermodynamics in recent times, comprehending their universal properties remains a difficult endeavor. By leveraging Duan's theory of $\varphi$-mapping topological currents \cite{qq6.1,qq6.2}, topological methods offer a fresh perspective. Wei et al. treat black hole states (solutions) as topological defects, and various black hole systems are sorted into three categories, each determined by the topological numbers linked to their respective topological charges \cite{qq7}. Based on this,  various types of black hole solutions have been studied as topological defects \cite{fm1,fm2,fm3,fm4,fm5,fm6,fm7,fm9,fm10,fm11,fm12,fm13,fm14,fm15,fm16,fm17,fm18,fm19,fm20,fm21}. They also introduced the topological method for studying black hole critical points, assigning each critical point a topological charge through a temperature-dependent function, they assigned topological charges to each critical point, classifying them into two types: the conventional critical
point and the novel critical point \cite{qq9}. Recent studies have widely investigated the topological classification of critical points across different black holes in gravitational backgrounds \cite{tp1,tp2,tp3,tp4,tp5,tp6,tp7,tp8,tp9,tp10}.
Following this, by investigating the asymptotic behavior of the vectors, they introduced a framework that categorized black hole states into four universal classifications, providing a clear system for classifying stable and unstable black hole configurations as the radius of the event horizon increases \cite{qq8}. For further examples of recent developments, see references \cite{uf1,uf2,uf3,uf4,uf5}. Building on this, Wu et al. found that black holes in the framework of gauge supergravity theory \cite{qq15,qq15.1}, and multiply rotating Kerr-AdS black holes \cite{qq14} exhibit an unusual thermodynamic stability at the low-to-high temperature limits of the Hawking temperature, which cannot be accounted for by the four existing topological classifications. This led to the expansion of the classification including five topological classes and three subclasses. In addition, topological methods have also been employed in the study of light rings \cite{lr1,lr2,lr3,lr4} and timelike circular orbits \cite{lr5,ws1,ws3}.

While significant advancements have been achieved in the topological classification of black holes, drawing significant global attention, an important and unresolved question remains: Are the five established topological categories and three subclasses sufficient? This study demonstrates that they are not. Through the study of the Ho\u{r}ava-Lifshitz (HL) black holes \cite{qq16,qq17}, we uncover a previously unrecognized topological subclass, named $\ddot{W}^{1-}$. This novel subclass showcases distinct black hole phases at high temperatures, with a stable large black hole and two unstable small black holes emerging in the phase space. As a result, our findings emphasize the need for an expansion of the current topological classification framework to include this newly identified subclass.

The structure of this paper is outlined as follows: In Section \ref{QQ}, we presents a brief overview of the thermodynamic topological methodology \cite{qq7} to set the stage for a direct comparison with the newly introduced topological subclass in Section \ref{KK}. There, we examine the general thermodynamic topological categories for charged black holes in the $z=3$ Ho\u{r}ava-Lifshitz gravity theory within the canonical ensemble. We introduce the novel $\ddot{W}^{1-}$ subclass and highlight its unique features through a detailed comparison in Section \ref{KK}. Finally, Section \ref{GG} contains our conclusions, where we reflect on the implications of our findings and suggest the potential for a new topological subclass, $\ddot{W}^{1+}$, to be explored in future work.
\section{A concise overview of the thermodynamic topological method, five topological classes and three subclasses}\label{QQ}
In this section, we offer a succinct overview of  the five established topological categories, along with their three associated subclasses. Based on this framework, a black hole is defined by its mass and entropy within a cavity \cite{qq7}, enabling the free energy $\mathcal{F}$ to be formulated as:
\begin{equation}
\label{n1}
\mathcal{F}=M-\frac{S}{\tau},
\end{equation}
\begin{table}[b]
\caption{Topological (sub)classes: The orientation of the $\phi^{r_h}$ arrows and the associated topological numbers across the four segments. \label{TabI}}
\begin{tabular}{c|c|c|c|c|c}\hline\hline
Topological (sub)classes     & $I_{1}$ &$I_{2}$ & $I_{3}$ & $I_{4}$ &$W$ \\ \hline
$W^{1-}$      & $\leftarrow$ & $\uparrow$& $\rightarrow$ & $\downarrow$ &-1 \\
$W^{0+}$   & $\leftarrow$   &$\uparrow$ & $\leftarrow$     &$\downarrow$ &0 \\
$W^{0-}$   & $\rightarrow$ &$\uparrow$ & $\rightarrow$ & $\downarrow$ &0 \\
$W^{1+}, ~\overline{W}^{1+}, ~\hat{W}^{1+},\widetilde{W}^{1+}$      &  $\rightarrow$   & $\uparrow$& $\leftarrow$ & $\downarrow$ &+1\\
$W^{0-\leftrightarrow 1+}$ (when $W = 0$)  & $\rightarrow$ &$\uparrow$ & $\rightarrow$ & $\downarrow$ &0 \\
$W^{0-\leftrightarrow 1+}$ (when $W = 1$) &  $\rightarrow$   & $\uparrow$& $\leftarrow$ & $\downarrow$ &+1 \\
\hline\hline
\end{tabular}
\end{table}
where, $\tau$  denotes the inverse temperature parameter, the free energy reduces to an on-shell quantity only when $\tau=\beta = 1/T$, as detailed in \cite{qq8}. With this relation, Equation~(\ref{n1}) reduces to
\begin{equation}
F=M-TS.
\end{equation}
To provide a detailed investigation of the topological properties, an extra parameter $\Theta$ is introduced, which varies within the range $(0, \pi)$. This allows for the construction of a two-component vector field \cite{qq7}, which can be expressed as
\begin{equation}\label{GS1}
\phi=\left(\phi^{r_h}, \phi^{\Theta}\right)=\left(\frac{\partial \tilde{\mathcal{F}}}{\partial r_h}, \frac{\partial \tilde{\mathcal{F}}}{\partial \Theta}\right),
\end{equation}
where, the function $\tilde{\mathcal{F}}$ is introduced as
\begin{equation}
\tilde{\mathcal{F}}=\mathcal{F}+\frac{1}{\sin \Theta}.
\end{equation}
The significance lies in the radial component of the vector field $(\phi^{r_{h}})$ being zero, as this allows the black hole state to be associated with the zero point. Based on the mapping topological current theory \cite{qq6.2}, the topological current is well-defined, with its explicit expression given as follows:
\begin{equation}
j^\mu=\frac{1}{2 \pi} \varepsilon^{\mu v \rho} \varepsilon_{a b} \partial_v n^a \partial_\rho n^b, \quad \mu, v, \rho=0,1,2 .
\end{equation}
Given that the topological current $j^\mu$ meets the conservation condition $(\partial_\mu j^\mu = 0)$, it can be written in terms of the Jacobian determinant
\begin{equation}
j^\mu=\delta^2(\phi) J^\mu\left(\frac{\phi}{x}\right).
\end{equation}
Owing to the nature of the function $\delta(\phi)$,  it is determined that the topological current vanishes everywhere except at isolated points. The total topological number, obtained from integrating the time component across the full parameter space, equals the sum of the local winding numbers at the zero points of each black hole
\begin{equation}
W=\int_{\Sigma} j^0 d^2 x=\sum_{i=1}^N \beta_i \eta_i=\sum_{i=1}^N w_i .
\end{equation}
In this context, $\beta_i$ denotes the Hopf index, while the Brouwer degree $(\eta_i)$ is used to define the orientation of the zero-point mapping. The winding number at the zero point of the black hole results from the product of these two factors $(\beta_i,\eta_i)$, as determined by the closed curve. For each zero point, a positive winding number represents a locally stable configuration, whereas a negative value reflects instability. The total topological number is obtained by taking the algebraic sum of all winding numbers. This theoretical approach enables the development of a systematic classification for different types of black holes.
To begin, we offer a succinct summary of the five well-established topological classifications and their three related subclasses, as described in \cite{qq9,qq14,qq15}
\begin{equation}\label{ch1}
W^{1-}, W^{0+}, W^{0-}, W^{1+},  \bar{W}^{1+}, \hat{W}^{1+},\widetilde{W}^{1+}, W^{0-\leftrightarrow 1+}.
\end{equation}
The asymptotic behaviors of the Hawking temperatures corresponding to these distinct topological classes or subclasses, focusing on the limits where the event horizon radius $(r_h)$ approaches the minimum radius and extends toward infinity for the black hole, are summarized as follows
\begin{equation}\label{ch2}
W^{1-} \quad: \beta\left(r_m\right)=0, \quad \beta(\infty)=\infty,
\end{equation}
\begin{equation}\label{ch3}
W^{0+} \quad: \beta\left(r_m\right)=\infty, \quad \beta(\infty)=\infty,
\end{equation}
\begin{equation}\label{ch4}
W^{0-}, \widetilde{W}^{1+} \quad: \beta\left(r_m\right)=0, \quad \beta(\infty)=0,
\end{equation}
\begin{equation}\label{ch5}
W^{1+}, \hat{W}^{1+} \quad: \beta\left(r_m\right)=\infty, \quad \beta(\infty)=0,
\end{equation}
\begin{equation}\label{ch6}
W^{0-\leftrightarrow 1+}, \bar{W}^{1+}: \beta\left(r_m\right)=\text { fixed temperature }, \beta(\infty)=0.
\end{equation}
In this context, $r_m$ represents the minimum horizon radius of the black hole, which may either be zero or nonzero. For example, in the case of the Reissner-Nordstr\"om black hole, $r_m$ corresponds to the magnitude of the black hole's mass or charge. In contrast, for the Schwarzschild black hole, $r_m$ is zero.
\begin{widetext}
\begin{table}[H]
\centering
\begin{tabular}{c|p{1.5cm}<{\centering}|p{1.5cm}<{\centering}|p{1.5cm}<{\centering}|p{3cm}<{\centering}|p{2cm}<{\centering}|p{1.5cm}<{\centering}}
 \hline \hline
Topological (sub)classes  & Innermost & Outermost & Low $T$ ($\beta\to\infty$) & High $T$ ($\beta\to 0$) & DP & $W$ \\ \hline
$W^{1-}$     & unstable & unstable & unstable large & unstable small  & in pairs & $-1$ \\ \hline
$W^{0+}$    & stable     & unstable & unstable large + stable small    & no & one more GP & $0$ \\ \hline
$W^{0-}$     & unstable & stable     & no & unstable small + stable large  & one more AP & $0$ \\ \hline
$W^{1+}$    & stable     & stable     & stable small  & stable large & in pairs & $+1$\\ \hline
$W^{0-\leftrightarrow 1+}$     & unstable & stable & no & stable large  & one more AP & $0$ or $+1$ \\ \hline
$\overline{W}^{1+}$    & stable  & stable   & no & stable large & in pairs & $+1$ \\ \hline
$\hat{W}^{1+}$    & stable     & stable     & unstable small+two stable small  & stable large & one more GP & $+1$\\ \hline
$\widetilde{W}^{1+}$   & unstable     & stable     & stable small& unstable small+stable small+stable large & one more AP & $+1$ \\
\hline \hline
\end{tabular}
\caption{The thermodynamic behavior of black holes in the eight topological (sub)classes-$W^{1-}$, $W^{0+}$, $W^{0-}$, $W^{1+}$, $W^{0-\leftrightarrow 1+}$, $\overline{W}^{1+}$, $\hat{W}^{1+}$, and $\widetilde{W}^{1+}$-is detailed in the table below.\label{TabII}}
\end{table}
\end{widetext}
Next, we analyze the asymptotic behavior of the vector fields near the boundary, as outlined in Eqs. (\ref{ch2})–(\ref{ch6}); the boundary in this context is defined by the contour $C = I_1 \cup I_2 \cup I_3 \cup I_4$, where
\begin{equation}
\begin{aligned}
& I_1=\left\{r_h=\infty, \Theta \in(0, \pi)\right\}, \\
& I_2=\left\{r_h \in\left(\infty, r_m\right), \Theta=\pi\right\}, \\
& I_3=\left\{r_h=r_m, \Theta \in(\pi, 0)\right\}, \\
& I_4=\left\{r_h \in\left(r_m, \infty\right), \Theta=0\right\} .
\end{aligned}
\end{equation}
The contour in question encloses every conceivable parameter region. With $\phi$ defined as orthogonal to $I_2$ and $I_4$ \cite{qq9}, we focus our analysis on its asymptotic behavior along $I_1$ and $I_3$. We begin by considering the $r_h$ component, which we express through the first law as
\begin{equation}
\phi^{r_h}=\frac{\partial \tilde{\mathcal{F}}}{\partial r_h}=\frac{\partial S}{\partial r_h}\left(\frac{1}{\beta}-\frac{1}{\tau}\right).
\end{equation}
Given that the cavity temperature $\tau$ is a constant positive value, for $\frac{\partial S}{\partial r_{h}} > 0$, the behavior of $\phi^{r_{h}}$ depends solely on $\beta$: it becomes positive as $\beta \to 0$ and negative when $\beta \to \infty$. Consequently, near the boundaries where $r_h \to r_m$ and $r_h \to \infty$, the vector $\phi$ points either to the right or left, with its direction determined by $\phi^{\Theta}$.
\begin{figure}[h]
\begin{center}
\includegraphics[width=0.45\textwidth]{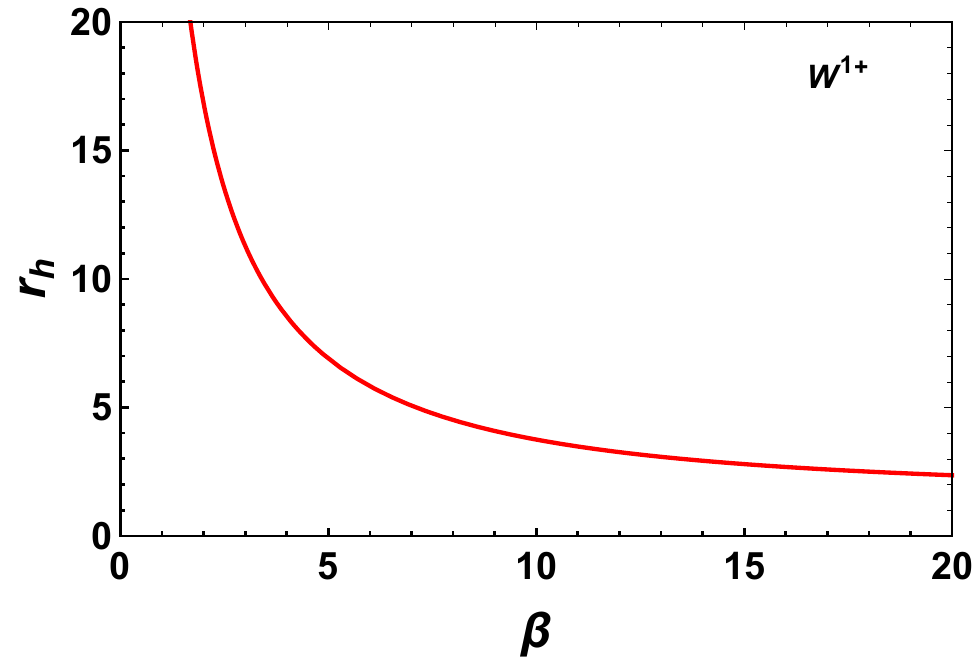}
\caption{In the $r_h-\beta$ plane, the zero points of the vector $\phi^{r_h}$ for the charged HL black hole, with the parameters $k / r_0 = 1$, $q / r_0 = 1$, and $P r_0^2 = 0.01$. The stable branch is represented by the red curve.}
\label{meng11}	
\end{center}
\end{figure}
In Table \ref{TabI}, the directions of the $\phi^{r_h}$ arrows for the four segments of each of the eight topological (sub)classes defined in Eq. (\ref{ch1}) are summarized, together with their corresponding topological numbers. Table \ref{TabII} outlines the five topological categories and three subcategories of black holes, separating the smallest (innermost) and largest (outermost) states, while also addressing their stability features under both low and high Hawking temperature regimes.
\section{The newly defined topological subclass: $\ddot{W}^{1-}$} \label{KK}
Considering a general dynamical coupling constant $\lambda$, Cai et al. derived the solution for topological black holes in HL gravity \cite{qq17,qq18}, which is expressed by the metric
\begin{equation}
d s^2=-N^2(r) f(r) d t^2+\frac{d r^2}{f(r)}+r^2 d \Omega_k^2.
\end{equation}
In this case, $d \Omega_k^2$ refers the metric element in a two-dimensional space, characterized by a constant scalar curvature of $2 k$. To cover all cases, we choose $k=-1,0,1$, representing the hyperbolic, flat, and spherical horizons, respectively. In the HL gravitational background, the solution for the charged black hole under the influence of an external electromagnetic field is derived for $s=1/2$ and $\lambda=1$, with the following metric functions given by
\begin{equation}
N(x)=1,
\end{equation}
\begin{equation}
f(x)=k+\frac{x^2}{1-\epsilon^2}-\frac{\sqrt{\epsilon^2 x^4+\left(1-\epsilon^2\right)\left(c_0 x-q^2 / 2\right)}}{1-\epsilon^2}.
\end{equation}
In the limit $\epsilon \rightarrow 1$, we obtain the AdS Reissner-Nordstr\"om black hole. Our primary focus is on the case where $\epsilon^2 = 0$. Setting $\epsilon^2 = 0$, the function \cite{qq16} becomes
\begin{equation}
f(x)=k+x^2-\sqrt{c_0 x-\frac{q^2}{2}},
\end{equation}
with $q$ and $c_0$ being constants of integration. The constant $c_0$ can be written as $c_0 = \frac{2 k^2 + q^2 + 4 k x_{+}^2 + 2 x_{+}^4}{2 x_{+}}$, and $x_{+} = \sqrt{-\Lambda} r_h$ serves as a root of $f(x_{+}) = 0$, where $r_h$ represents the radius of the event horizon. The pressure is tightly coupled to the cosmological constant $\Lambda$ in the extended phase space, and is given by $P = -\frac{\Lambda}{8 \pi}$ \cite{mjl1}. From this, the mass and charge of the black holes are derived as
\begin{equation}\label{gg1}
M=\frac{c^3\left(2 k^2+q^2+32 k \pi P  r_h^2+128 P^2 \pi^2 r_h^4\right) \Omega_k}{256 \pi^2 r_hG P},
\end{equation}
\begin{equation}
Q=\frac{q \Omega_k c^3}{32 \sqrt{2} \pi^{3 / 2} G \sqrt{P}}.
\end{equation}
\begin{figure}[h]
\begin{center}
\includegraphics[width=0.45\textwidth]{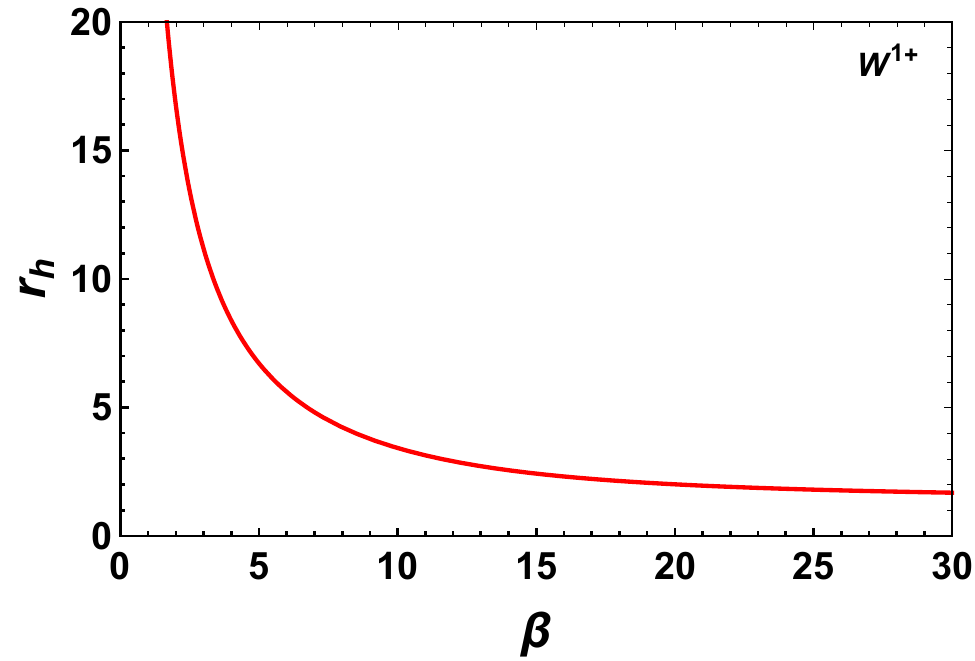}
\caption{In the $r_h-\beta$ plane, the zero points of the vector $\phi^{r_h}$ for the charged HL black hole, using parameters $k / r_0 = 0$, $q / r_0 = 1$, and $P r_0^2 = 0.01$. The stable branch is represented by the red curve.}
\label{meng12}	
\end{center}
\end{figure}
\begin{figure}[h]
\begin{center}
\includegraphics[width=0.45\textwidth]{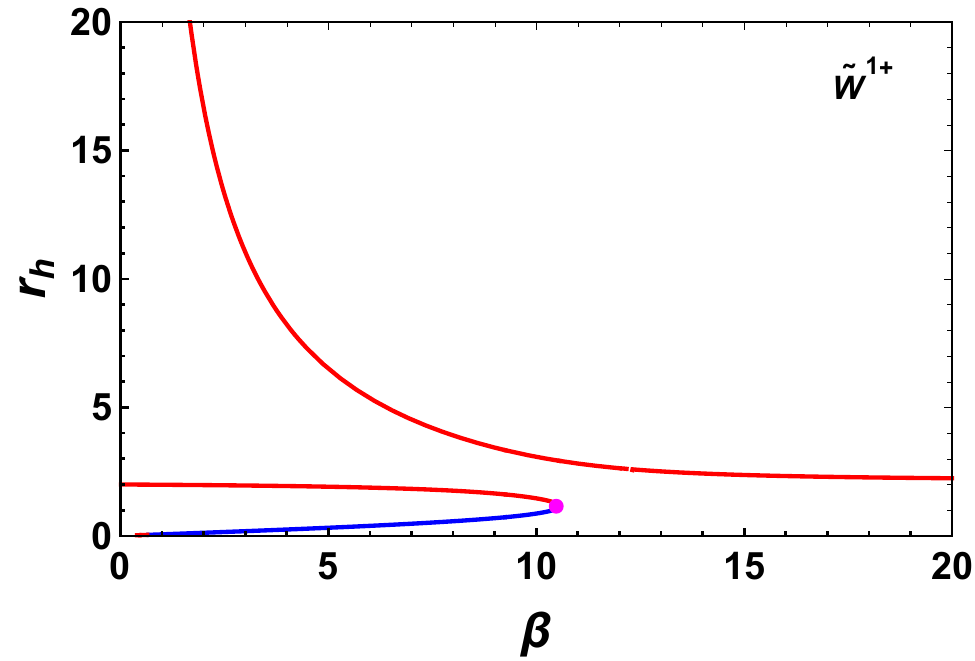}
\caption{In the $r_h-\beta$ plane, the zero points of the vector $\phi^{r_h}$ for the charged HL black hole, with parameters $k / r_0 = -1$, $q / r_0 = 1$, and $P r_0^2 = 0.01$. The blue unstable branch ($w = -1$) and the red stable branch ($w = +1$) meet at the annihilation point (AP) marked by the pink dot. The count of stable and unstable states yields two stable and one unstable state, leading to a total topological number of $W = -1 + 1 + 1 = 1$.}
\label{meng13}	
\end{center}
\end{figure}
The thermodynamic quantities, including the temperature, entropy, and electromagnetic potential are given by
\begin{equation}
T=\frac{32 k^2 P \pi r_h^2+384 P^2 \pi^2 r_h^4-2 k^2-q^2}{16 k \pi r_h+128 \pi^2 P r_h^3},
\end{equation}
\begin{equation}\label{gg2}
S=\frac{c^3 \Omega_k\left[k \operatorname{In}\left(2 \sqrt{2 \pi P} r_h\right)+4 P \pi r_h^2\right]}{16 G P \pi}+S_0,
\end{equation}
\begin{equation}
\Phi=\frac{q}{2 \sqrt{2 \pi P} r_h}+\Phi_0.
\end{equation}
In this context, $S_0$ and $\Phi_0$ are constant values. The following analysis delves into the universal classes of thermodynamic topology for this black hole within the framework of canonical and grand canonical ensembles, respectively.
\begin{figure}[h]
\begin{center}
\includegraphics[width=0.45\textwidth]{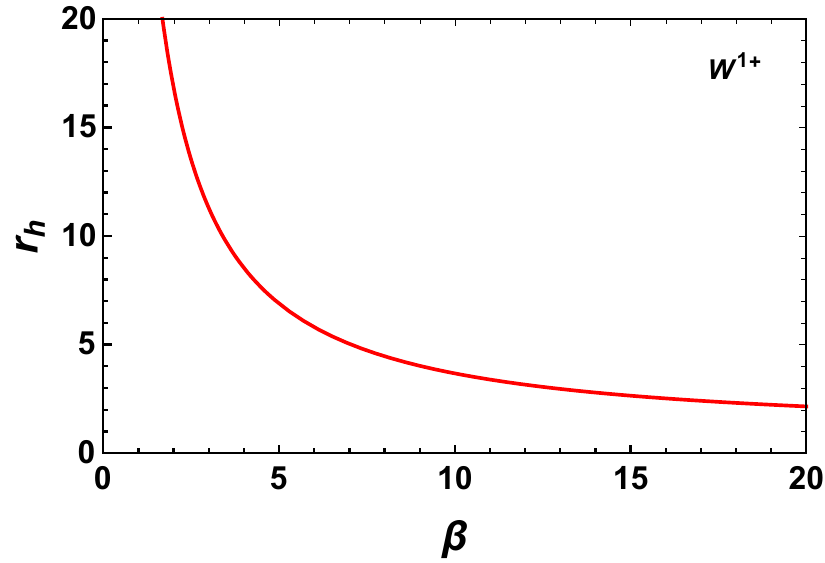}
\caption{In the $r_h-\beta$ plane, the zero points of the vector $\phi^{r_h}$ for the charged HL black hole, with the parameters $k / r_0 = 1$, $q / r_0 = 1$, and $P r_0^2 = 0.01$. The stable branch is represented by the red curve.}
\label{meng21}	
\end{center}
\end{figure}
\begin{figure}[h]
\begin{center}
\includegraphics[width=0.45\textwidth]{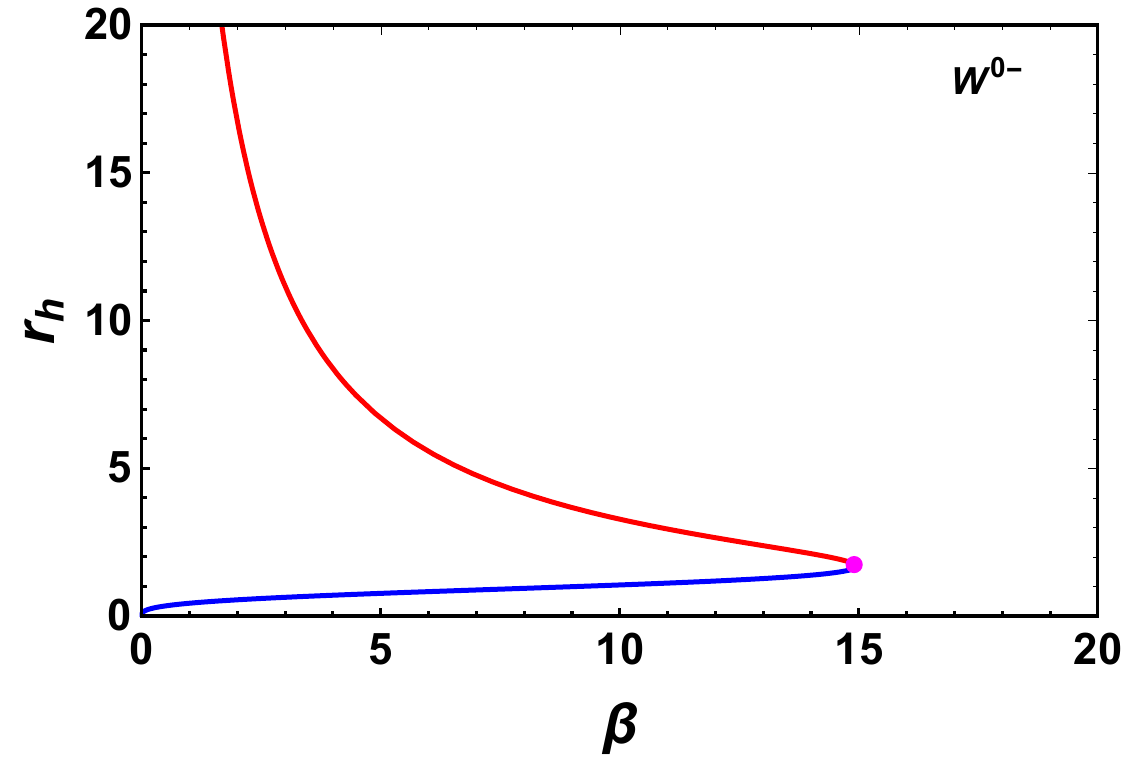}
\caption{In the $r_h-\beta$ plane, the zero points of the vector $\phi^{r_h}$ for the charged HL black hole, with parameters $k / r_0=0$, $q / r_0=1$, and $P r_0^2=0.01$. At the intersection of the blue unstable ($w=-1$) and red stable ($w=+1$) branches, denoted by the pink annihilation point (AP).}
\label{meng22}	
\end{center}
\end{figure}

\subsection{Black holes in the context of the canonical ensemble}\label{IIII}
In this subsection, we will examine the the asymptotic behavior of charged black holes with different horizons (spherical, flat and hyperbolic) in the context of the canonical ensemble, which enables the topological classification of black holes. Under this ensemble, the   volume, Hawking temperature, and particle number are held constant. By inserting equation (\ref{gg1}) and equation (\ref{gg2}) into equation (\ref{n1}), we can derive
\begin{equation}
\begin{aligned}
\mathcal{F} & =M-\frac{S}{\tau}\\
& =\frac{c^3 \Omega_{k}\left(2 k^2+q^2+32 kP \pi r_{h}^2+128 P^2 \pi^2 r_{h}^4\right)}{256 GP \pi^2 r_{h}} \\
& -\frac{c^3\left(S_0+4 kIn\left[2 \sqrt{2 \pi P} r_{h}\right]+16 \pi P r_{h}^2\right) \Omega_{k}}{64 \pi \tau GP}.
\end{aligned}
\end{equation}
As a result, the elements of the vector $\phi$ are
\begin{equation}
\begin{aligned}
\phi^{r_{h}}& =\frac{c^3\left(384 P^2 \pi^2 r_h^4-2 k^2-q^2-32 k P \pi r_h^2\right) \Omega_k}{512 \sqrt{2} G P^{3 / 2} \pi^{5 / 2} r_h^2} \\
& -\frac{c^3\left(k+8 P \pi r_h^2\right) \Omega_k}{32 \sqrt{2} G P^{3 / 2} \pi^{3 / 2} \tau r_h},
\end{aligned}
\end{equation}
\begin{equation}
\phi^{\Theta}=-\cot \Theta \csc \Theta.
\end{equation}
By considering the condition that $\phi^{r_h} = 0$, the inversion temperature parameter $\tau$ can be represented by
\begin{equation}
\tau=\beta=\frac{16 \pi r\left(k+8 P \pi r^2\right)}{32 k P \pi r^2+384 P^2 \pi^2 r^4-2 k^2-q^2}.
\end{equation}
We now proceed to investigate the universal thermodynamic topological classification for varying black hole parameters. In this work, we consider parameter values $c = G = \Omega_k = q = 1$.\\
$(i)$ Spherical and flat horizons:\\
The asymptotic behavior of the parameter $\beta$ is found to be fully consistent with equation (\ref{ch5}). As depicted in Figure (\ref{meng11}), the radius of the event horizon steadily diminishes as the inverse temperature parameter $(\beta)$ grows, indicating that only a single stable black hole exists for all values of $\beta$, with no phase transition observed. Additionally, the black hole exhibits stability in both the smallest and largest radius regions; at low and high temperatures, these correspond to stable small- and large-mass black holes, respectively. According to the universal thermodynamic classification presented in \cite{qq8}, the black hole with a spherical event horizon is classified as $W^{1+}$ category. In the case where $k=0$, the event horizon takes on a flat structure. The relationship between the inverse temperature parameter and the radius of the event horizon is depicted in Figure (\ref{meng12}). Black holes remain stable across the entire temperature spectrum, and those with a flat event horizon are classified as belonging to the $W^{1+}$ category.
\\$(ii)$ The hyperbolic horizon:\\
As shown in Figure (\ref{meng13}), as $\beta \to \infty$ (low temperature limit), black holes with hyperbolic horizons remain in a stable small black hole phase. In contrast, as $\beta \to 0$ (high temperature limit), the system reveals a total of three black hole phases, which include both stable and unstable small black hole states, as well as a large black hole that maintains stability. The two smaller black hole phases disappear at the annihilation point marked in pink ($\beta_c = 10.55$). According to the thermodynamic classification framework described in \cite{qq14}, these black holes are categorized as part of the topological subclass $\widetilde{W}^{1+}$.
\begin{figure}[h]
\begin{center}
\includegraphics[width=0.45\textwidth]{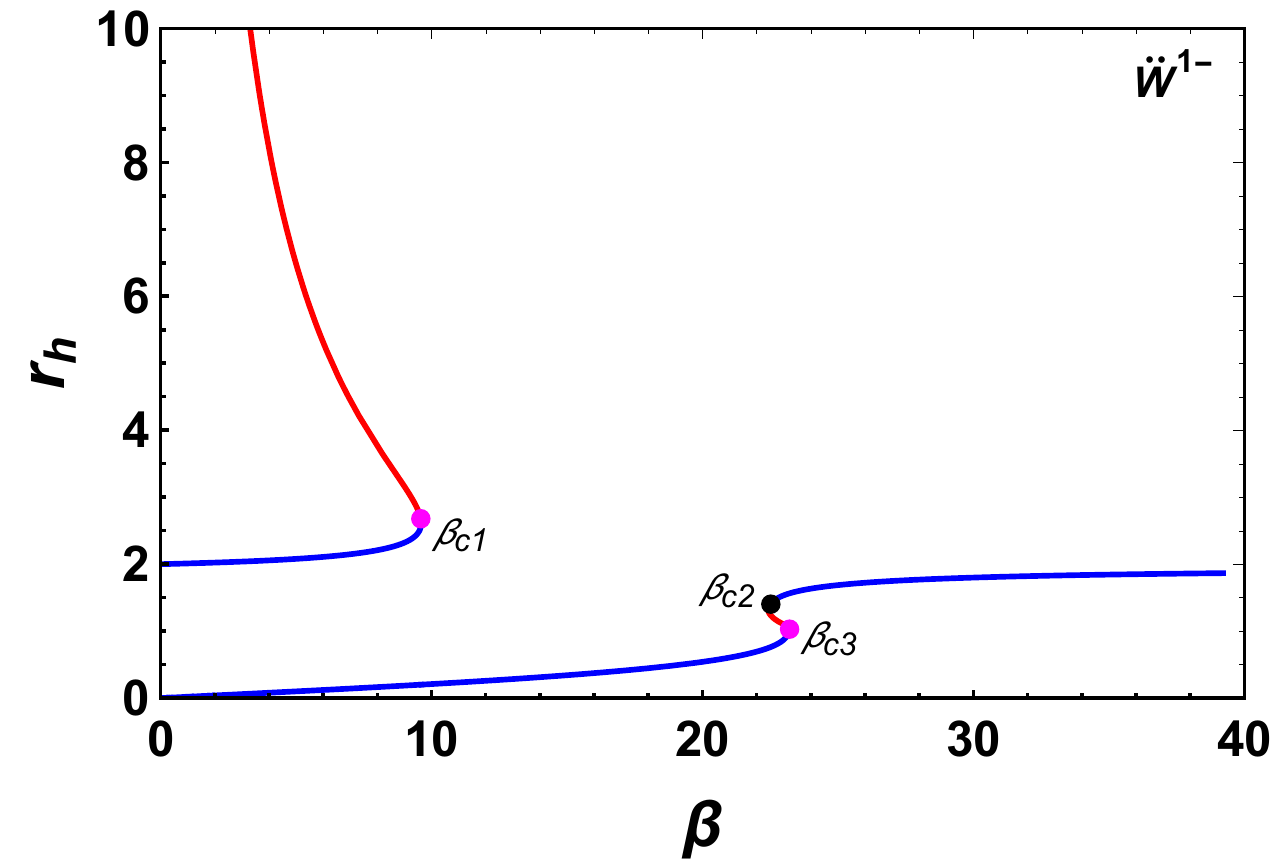}
\caption{In the $r_h-\beta$ plane, the zero points of the vector $\phi^{r_h}$ for the charged HL black hole, with parameters $k / r_0=-1$, $q / r_0=1$, and $P r_0^2=0.01$. The annihilation point (AP) is marked by a pink dot, while the generation point (GP) is indicated by a black dot. Under the condition $\beta<\beta_{c1}=9.61788$, the system exhibits two unstable small black hole states and one stable large black hole state, resulting in the total topological number $W=-1-1+1=-1$. For the temperature range $\beta_{c2}=23.2166<\beta<\beta_{c3}=23.2196$, the system consists of two unstable states and one stable small black hole, yielding the total topological number $W=-1+1-1=-1$.}
\label{meng23}	
\end{center}
\end{figure}

\begin{figure}[h]
\begin{center}
\includegraphics[width=0.45\textwidth]{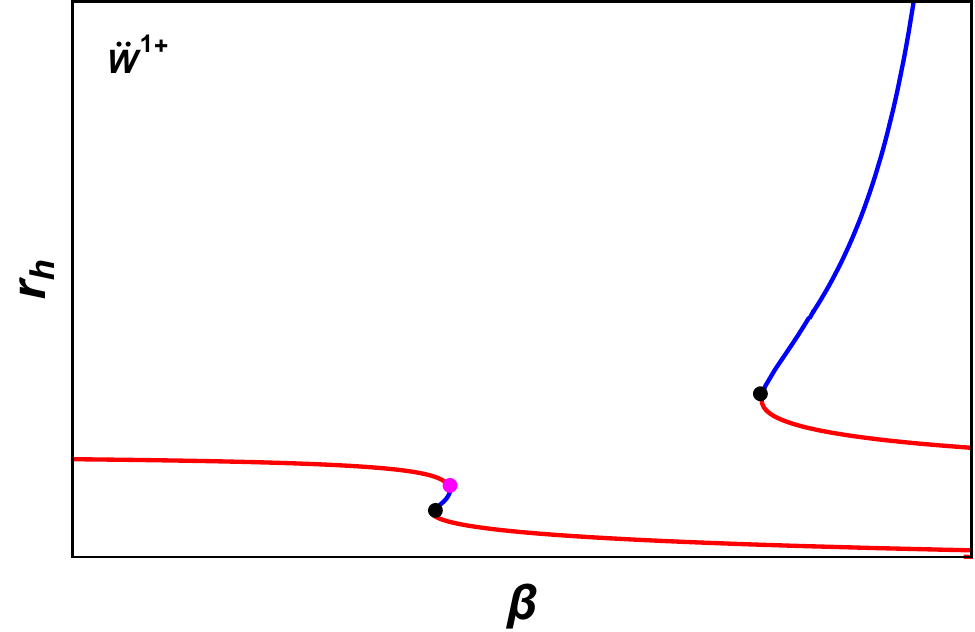}
\caption{The schematic representation of the zero points of $\phi^{r_h}$ is displayed on the $r_h-\beta$ plane, illustrating the potential new topological subclass $(\ddot{W}^{1+})$.}
\label{menghy}	
\end{center}
\end{figure}
\subsection{Black holes in the context of the grand canonical ensemble }\label{IIIII}
In this subsection, we examine the asymptotic behavior of black holes with various types of horizons (spherical, flat, and hyperbolic) under the grand canonical ensemble. Here, the system can exchange both energy and charge with the external environment, while its temperature, volume, and chemical potential are constrained. Within this framework, the generalized free energy $(\mathcal{F})$ is defined as
\begin{equation}
\begin{aligned}
\mathcal{F}& =M-Q \Phi-\frac{S}{\tau} \\
& =\frac{c^3 \Omega_k\left(32 P^{3 / 2} r_h^3+\frac{\sqrt{2} q_0}{\pi^{3 / 2}}\right)}{64 G \sqrt{P}} \\
& +\frac{c^3 \Omega_k\left(2 k^2-q^2+32 k P \pi r_h^2\right)}{256 G P \pi^2 r_h} \\
& -\frac{16 \pi G P S_0+c^3 \Omega_k\left[k \operatorname{In}\left(2 \sqrt{2 \pi P} r_h\right)+4 P \pi r_h^2\right]}{16 \pi \tau G P}.
\end{aligned}
\end{equation}
In this situation, the elements of the vector $\phi$ are given by
\begin{equation}
\begin{aligned}
\phi^{r_h} & =\frac{c^3\left(-2 k^2+q^2+32 k P \pi r_h^2+384 P^2 \pi^2 r_h^4\right) \Omega_k}{512 \sqrt{2} G P^{3 / 2} \pi^{5 / 2} r_h^2} \\
& -\frac{c^3 \Omega_k\left(k+8 P \pi r_h^2\right)}{32 \sqrt{2} G P^{3 / 2} \pi^{3 / 2} \tau r_h},
\end{aligned}
\end{equation}
and
\begin{equation}
\phi^{\Theta}=-\cot \Theta \csc \Theta.
\end{equation}
By imposing the condition $\phi^{r h} = 0$, it becomes possible to determine
\begin{equation}
\tau=\beta=\frac{16 \pi r_h\left(k+8 P \pi r_h^2\right)}{q^2+32 k P \pi r_h^2+384 P^2 \pi^2 r_h^4-2 k^2}.
\end{equation}
Next, we explore the universal classification of the charged HL black hole in the context of the grand canonical ensemble.\\
$(i)$ The spherical horizon:\\
As shown in Figure (\ref{meng21}), the black hole with a spherical horizon exhibits a single stable branch. The system remains thermodynamically stable at both low and high temperature limits, indicating that the topological structure of the black hole is preserved within the ensemble, thus classifying it within the $W^{1+}$ category.\\
$(ii)$ The flat horizon:\\
As depicted in Figure (\ref{meng22}), the thermodynamic phase diagram consists of a stable large black hole branch (represented by the red curve) and an unstable small black hole branch (indicated by the blue curve), with the total topological number $W = 0$. When the inverse temperature parameter exceeds the critical value $(\beta_{c} = 14.944)$-specifically at the pink annihilation point-the system can no longer support a black hole phase. In the high-temperature regime, both a stable large black hole and an unstable small black hole coexist within the system. According to the universal thermodynamic topological classification introduced in Reference \cite{qq8}, black holes with planar event horizons are classified under Class $W^{0-}$.
\\$(iii)$ The hyperbolic horizon:\\
In Figure (\ref{meng23}), two independent curves are shown. In the first curve located in the upper left area, two branches of black holes are observed.  We find that the asymptotic behavior of the inverse temperature parameter  coincides with that of the
$W^{1-}$ class, which is described as follows
\begin{equation}
\beta\left(r_m\right)=0, \quad \beta(\infty)=\infty.
\end{equation}
It can also be found that with the increase of the horizon radius, the black holes undergo a sequence of unstable $(\omega=-1)$ and stable $(\omega=1)$ phases.  The two states meet and ultimately vanish at the critical point, referred to as the annihilation point, where $\beta_{c1} = 9.61788$. The second curve reveals that within a particular temperature range, here are three branches of small black holes, with their stability alternating between unstable $(\omega=-1)$, stable $(\omega=1)$, and unstable $(\omega=-1)$ states as the radius of the event horizon increases. When the inverse temperature parameter satisfies $\beta < \beta_{c1} = 9.61788$, three black hole states coexist, leading to a total topological number of $W =-1-1+1=-1$. The system's asymptotic behavior differs from the eight previously established thermodynamic topological structures. In this case, the inner region of the black hole is unstable, whereas the outer region remains stable.
In the low-temperature limit, the system features a single unstable small black hole state. In the high-temperature limit, the system supports two unstable small black hole states along with one stable large black hole state. As the inverse temperature increases, the system shifts to a state with only one unstable small black hole. For temperatures within the range $\beta_{c2} = 23.2166 < \beta < \beta_{c3} = 23.2196$, the system contains two unstable small black hole branches and one stable large black hole branch, yielding a total topological number of $W  = -1$. Notably, throughout all values $(\beta)$, the topological number remains consistently $-1$, and the system's topological classification matches that of the $W^{1-}$ classes. As a result, we suggest that this system belongs to a new subclass, $\ddot{W}^{1-}$, within the topological classification framework. Furthermore, Table \ref{TabC}  provides a summary of the asymptotic behavior of the vector field near the boundaries $I_i$.

In this analysis, we explore the asymptotic behavior of the newly discovered topological subclass. Initially, the system consists of at least three black hole states: with an increasing event horizon radius, the system transitions through three phases: two unstable states $(\omega=-1)$ and one stable state $(\omega=1)$. In these phases, the heat capacities are negative for the first two states and positive for the final, stable state. Subsequent states appear in pairs, where each pair includes one state with a winding number of $-1$ and one with $+1$. The arrangement of winding numbers does not follow a simple alternating pattern but instead begins with $[-,+]$. This ensures that the innermost black hole is associated with $\omega = -1$ (an unstable state), and the outermost one with $\omega = +1$ (a stable state), forming a topological classification represented by $[-,+]$.
\section{Conclusions}\label{GG}
This study offers a comprehensive examination of the universal  classification of thermodynamic topology for charged black holes within the framework of Ho\u{r}ava-Lifshitz gravity. It identifies a new thermodynamic feature and introduces a novel topological subclass, denoted $(\ddot{W}^{1-})$. The detailed analysis yields several key conclusions, which are summarized below:\\
$(i)$ Charged black holes with spherical horizons display similar thermodynamic stability in both the canonical and grand canonical ensembles. In each ensemble, only a single stable branch exists. This behavior leads to their classification as type $W^{1+}$. In contrast, charged black holes with flat  horizon exhibit different stability characteristics depending on the ensemble. In the canonical ensemble, their thermodynamic behavior is comparable to that of spherically symmetric black holes. However, in another ensemble, no black hole solutions exist at low temperatures. At high temperatures, the small black hole is unstable, while the large black hole remains stable. As a result, they are classified as type $W^{0-}$.\\
$(ii)$ In the canonical ensemble, charged black holes with hyperbolic horizon exhibit a triphasic coexistence-comprising unstable small black holes, stable small black holes, and stable large black holes-within a specific range of temperatures. This behavior classifies them into the $\widetilde{W}^{1+}$ category. In another ensemble, the two distinct curves shown in Figure (\ref{meng23}) demonstrate the differing thermodynamic stability properties of these black holes. From this analysis, a new topological subclass, labeled $(\ddot{W}^{1-})$, is introduced, with its key characteristics outlined in Table \ref{Tab1}. This study highlights the increased complexity in black hole thermodynamics beyond prior models and underscores the importance of topological methods in revealing previously overlooked structural and dynamic aspects.\\
$(iii)$ As depicted in Figure (\ref{menghy}), the symmetry observed in the function $(\beta, r_h)$ suggests an intriguing possibility: the potential existence of a new subclass within the realm of thermodynamic topology. Table \ref{Tab2} shows that the stability properties of the $(\ddot{W}^{1+})$ subclass are directly opposite to those of the $(\ddot{W}^{1-})$ subclass. This intriguing distinction calls for further verification through detailed physical black hole solutions in future research, marking an important avenue for further exploration. The introduction of the $(\ddot{W}^{1-})$ subclass, along with its potential relationship to the $(\ddot{W}^{1+})$ subclass, highlights the need to expand the current topological classification framework to encompass emerging features not yet fully captured. These results offer a fresh theoretical basis for classifying thermodynamic systems and suggest the possible existence of additional, yet-to-be-identified topological categories.
\begin{table}[b]
\caption{New topological subclasses: The orientation of the $\phi^{r_h}$ arrows and the associated topological numbers across the four segments. \label{TabC}}
\begin{tabular}{c|c|c|c|c|c}\hline\hline
Topological (sub)classes     & $I_{1}$ &$I_{2}$ & $I_{3}$ & $I_{4}$ &$W$ \\ \hline
$\ddot{W}^{1-}$      & $\leftarrow$ & $\uparrow$& $\rightarrow$ & $\downarrow$ &-1 \\
\hline\hline
\end{tabular}
\end{table}
\begin{widetext}
\begin{table}[H]
\centering
\begin{tabular}{c|p{1.5cm}<{\centering}|p{1.5cm}<{\centering}|p{1.5cm}<{\centering}|p{3cm}<{\centering}|p{2cm}<{\centering}|p{1.5cm}<{\centering}}
 \hline \hline
Topological (sub)classes  & Innermost & Outermost & Low $T$ ($\beta\to\infty$) & High $T$ ($\beta\to 0$) & DP & $W$ \\ \hline
$(\ddot{W}^{1-})$    & unstable & stable & unstable small & unstable small+unstable small+stable  large   & one more AP & $-1$ \\
\hline \hline
\end{tabular}
\caption{The thermodynamic behavior of black holes belonging to the novel topological subclass $(\ddot{W}^{1-})$. \label{Tab1}}
\end{table}
\end{widetext}
\begin{widetext}
\begin{table}[H]
\centering
\begin{tabular}{c|p{1.5cm}<{\centering}|p{1.5cm}<{\centering}|p{3cm}<{\centering}|p{1.5cm}<{\centering}|p{2cm}<{\centering}|p{1.5cm}<{\centering}}
 \hline \hline
Topological (sub)classes  & Innermost & Outermost & Low $T$ ($\beta\to\infty$) & High $T$ ($\beta\to 0$) & DP & $W$ \\ \hline
$(\ddot{W}^{1+})$     & stable & unstable &stable small+stable small+unstable large & stable small  & one more GP  & $+1$ \\
\hline \hline
\end{tabular}
\caption{Proposed thermodynamic behaviors of the black holes in the possible $(\ddot{W}^{1+})$ topological subclass. \label{Tab2}}
\end{table}
\end{widetext}
\section*{Acknowledgments}
\hspace{0.5cm}
We are grateful to Dr. Di Wu for useful discussions on the topological subclass. We also acknowledge the anonymous referees for their
valuable comments on improving our paper. This work is supported  by the National Natural Science Foundation of China (NSFC) under Grants No. 12265007, 12405081, 11865018.  by the Doctoral Foundation of Zunyi Normal University of China (BS [2022] 07). Also, the research of H.H. was supported by the Q-CAYLE project, funded by the European Union-Next Generation UE/MICIU/Plan de Recuperacion, Transformacion y Resiliencia/Junta de Castilla y Leon (PRTRC17.11), and also by project PID2023-148409NB-I00, funded by MICIU/AEI/10.13039/501100011033. Financial support of the Department of Education of the Junta de Castilla y Leon and FEDER Funds is also gratefully acknowledged (Reference: CLU-2023-1-05). Additionally, H. H. is grateful to Excellence project
FoS UHK 2203/2025-2026 for the financial support.

\end{document}